\documentclass[twocolumn,showkeys,aps,prb,showpacs]{revtex4-1}
\usepackage{graphicx}
\usepackage[CJKbookmarks,dvipdfm,colorlinks,linkcolor=blue,citecolor=blue]{hyperref}

\begin{document}

\title{Structure  effect on intrinsic piezoelectricity in septuple-atomic-layer  $\mathrm{MSi_2N_4}$ (M=Mo and W)}

\author{San-Dong Guo$^{1}$, Yu-Tong Zhu$^{1}$,  Wen-Qi Mu$^{1}$, Lei Wang$^{2,3}$  and  Xing-Qiu Chen$^{2,3}$}
\affiliation{$^1$School of Electronic Engineering, Xi'an University of Posts and Telecommunications, Xi'an 710121, China}
\affiliation{$^2$Shenyang National Laboratory for Materials Science, Institute of Metal Research,
Chinese Academy of Science, 110016 Shenyang, Liaoning, P. R. China}
\affiliation{$^3$School of Materials Science and Engineering, University of Science and Technology of China,
Shenyang 110016, P. R. China}

\begin{abstract}
The recently experimentally synthesized monolayer $\mathrm{MoSi_2N_4}$ and  $\mathrm{WSi_2N_4}$ (\textcolor[rgb]{0.00,0.00,1.00}{Science 369, 670-674 (2020})) lack inversion symmetry, which
allows them to become piezoelectric. In this work,  based on ab initio calculations, we report structure  effect on intrinsic piezoelectricity in septuple-atomic-layer  $\mathrm{MSi_2N_4}$ (M=Mo and W), and six structures ($\alpha_i$ ($i$=1 to 6)) are considered with  the same space group.
These structures can  connect to each other through   translation, mirror and rotation  operations of double layer unit $\mathrm{Si_2N_2}$. It is found that  $\mathrm{MSi_2N_4}$ (M=Mo and W)  with $\alpha_i$ ($i$=1 to 6) all are indirect band gap semiconductors.  Calculated results show that $\mathrm{MoSi_2N_4}$ and  $\mathrm{WSi_2N_4}$ monolayers have the same structural dependence on piezoelectric  strain and stress  coefficients ($d_{11}$ and $e_{11}$), together with the  ionic and  electronic contributions to  $e_{11}$. The   $\alpha_5$  phase  has  largest $d_{11}$ for both $\mathrm{MoSi_2N_4}$ and  $\mathrm{WSi_2N_4}$, which are  larger than 2.9 pm/V. Finally, we investigate the intrinsic piezoelectricity of  monolayer  $\mathrm{MA_2Z_4}$ (M=Cr, Mo and W; A=Si and Ge; Z=N and P)  with $\alpha_1$ and $\alpha_2$ phases expect  $\mathrm{CrGe_2N_4}$, because they all are semiconductors and their  enthalpies of formation between  $\alpha_1$ and  $\alpha_2$ phases are very close. The most important result is that  monolayer $\mathrm{MA_2Z_4}$ containing P atom have more stronger  piezoelectric polarization than one including N atom.
The largest $d_{11}$  among $\mathrm{MA_2N_4}$ materials is 1.85 pm/V, which is close to the smallest $d_{11}$ of  1.65 pm/V in $\mathrm{MA_2P_4}$ monolayers. For $\mathrm{MA_2P_4}$, the largest $d_{11}$ is up to 6.12 pm/V. Among the 22 monolayers,  $\alpha_1$-$\mathrm{CrSi_2P_4}$, $\alpha_1$-$\mathrm{MoSi_2P_4}$, $\alpha_1$-$\mathrm{CrGe_2P_4}$, $\alpha_1$-$\mathrm{MoGe_2P_4}$ and $\alpha_2$-$\mathrm{CrGe_2P_4}$
have   large $d_{11}$, which are  greater than or close to 5 pm/V, a typical
value for bulk piezoelectric materials.  These materials  are recommended for experimental exploration.
Our study reveals that the  $\mathrm{MA_2Z_4}$ family have  the potential  applications in  piezoelectric field.

\end{abstract}
\keywords{$\mathrm{MA_2Z_4}$ family, Piezoelectronics, 2D materials}

\pacs{71.20.-b, 77.65.-j, 72.15.Jf, 78.67.-n ~~~~~~~~~~~~~~~~~~~~~~~~~~~~~~~~~~~Email:sandongyuwang@163.com}

\maketitle

\section{Introduction}
In semiconductors or insulators with broken inversion symmetry, an intrinsic electromechanical
coupling between stresses and electric polarizations can be observed, which is called piezoelectric effect.
Two-dimensional (2D) materials can show  unique  properties compared to their bulk counterparts, and the reduction in dimensionality of 2D
materials can  often  eliminate inversion symmetry,  which allows these materials to become piezoelectric\cite{q4}.
It has been theoretically reported that many  2D
materials break inversion symmetry
and hence can exhibit piezoelectricity, such as  group IIA and IIB metal oxides, group-V binary semiconductors, transition metal dichalchogenides (TMD), Janus TMD and group III-V semiconductors\cite{q7,q7-2,q7-3,q7-3-1,q7-3-2,q9,q10,q11,q12,qr,nr,nr1}.  A majority of  structures have piezoelectric coefficients greater than  a typical value of  bulk piezoelectric materials (5 pm/V). Significantly,  the  monolayer SnSe,
SnS, GeSe and GeS  with puckered structure  possess   giant piezoelectricity,  as high as  75-251 pm/V\cite{q10}, which  may have huge potential application in the field of sensors, actuators and energy harvesters.
 The different crystal symmetry can induce  a only in-plane piezoelectricity like TMD monolayers\cite{q9}, both in-plane and out-of-plane piezoelectricity for example 2D  Janus monolayers\cite{q7,q7-3},  or a pure out-of-plane piezoelectricity such as penta-graphene\cite{q7-4}. It has been proved that strain may be a effective strategy to tune piezoelectric properties of 2D materials\cite{r1,r3}.
Experimentally discovered piezoelectricity of $\mathrm{MoS_2}$\cite{q5,q6}, MoSSe\cite{q8}  and $\mathrm{In_2Se_3}$\cite{q8-1}  has  triggered an intense interest in piezoelectric properties of  2D materials.

\begin{figure*}
  \includegraphics[width=16.0cm]{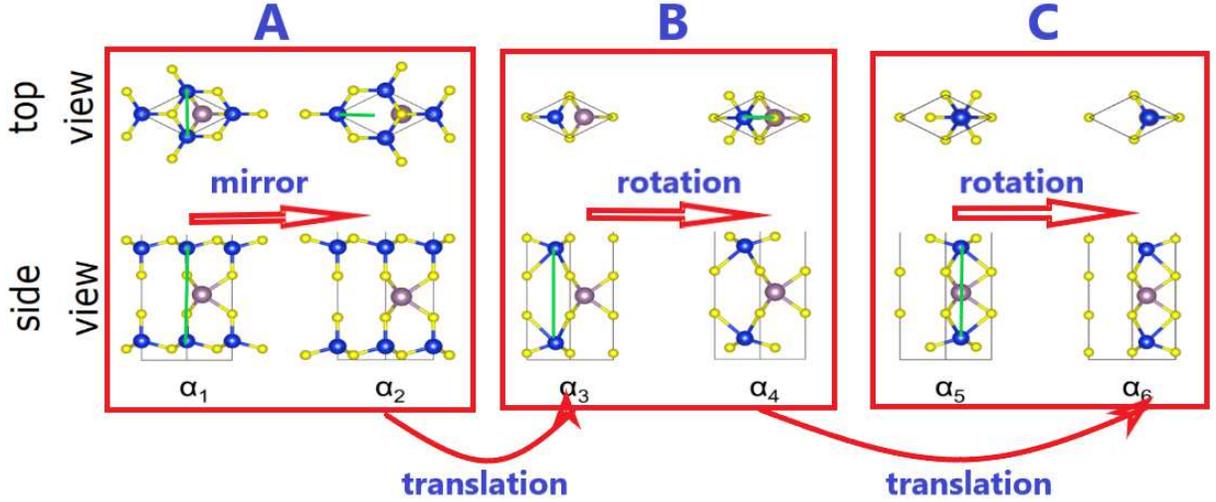}
  \caption{(Color online)The crystal structure of $\alpha_i$- ($i$=1 to 6) $\mathrm{MA_2Z_4}$  including  top  and  side views.   The purple balls represent M atoms, and the  blue balls for A atoms, and the yellow balls for Z atoms. These crystal structure  can be divided into three categories: A ($\alpha_1$, $\alpha_2$), B ($\alpha_3$, $\alpha_4$) and C ($\alpha_5$, $\alpha_6$) according to the relative positions of M and A atoms. The different categories can be connected by translation operation, and the different structures in the same category can be related by mirror or rotation operations. The green lines represent mirror face,  translation direction or rotation axis.}\label{t0}
\end{figure*}

It is  meaningful to  explore piezoelectricity of new 2D family. Recently, the layered
2D $\mathrm{MoSi_2N_4}$ has been  experimentally achieved by chemical vapor deposition (CVD)\cite{msn}, which possesses  semiconducting behavior, high strength and excellent ambient stability.  In rapid sequence, 2D
$\mathrm{WSi_2N_4}$ has also been synthesized by CVD. In the wake of $\mathrm{MSi_2N_4}$ (M=Mo and W), $\mathrm{MA_2Z_4}$ family are constructed with twelve different structures ($\alpha_i$ and  $\beta_i$ ($i$=1 to 6))  by intercalating $\mathrm{MoS_2}$-type  $\mathrm{MZ_2}$ monolayer into InSe-type  $\mathrm{A_2Z_2}$ monolayer\cite{m20}.  The  $\mathrm{MA_2Z_4}$ family spans a wide range of properties  from semiconductor to topological insulator to Ising superconductor upon the number of valence electrons (VEC).   Intrinsic piezoelectricity in monolayer  $\mathrm{XSi_2N_4}$ (X=Ti, Zr, Hf, Cr, Mo and W) with $\alpha_1$ phase  are studied  by the first principle calculations\cite{m21}, and the independent  in-plane piezoelectric constants $d_{11}$ is predicted to be 0.78 pm/V-1.24 pm/V. The valley-dependent properties of monolayer $\mathrm{MoSi_2N_4}$, $\mathrm{WSi_2N_4}$ and $\mathrm{MoSi_2As_4}$ have been investigated by the first-principle calculations\cite{g1}. The structural, mechanical, thermal, electronic, optical and photocatalytic properties of $\mathrm{MoSi_2N_4}$ are studied by using hybrid density functional theory (HSE06-DFT)\cite{g2}.

In this work, the role of crystal structure  on intrinsic piezoelectricity in monolayer  $\mathrm{MSi_2N_4}$ (M=Mo and W) are studied by using density functional perturbation theory (DFPT)\cite{pv6}. It is interesting to note that  the same structural dependence on $d_{11}$ and $e_{11}$  between monolayer $\mathrm{MoSi_2N_4}$ and $\mathrm{WSi_2N_4}$ is observed. Calculated results show  that the atomic arrangement of $\mathrm{A_2Z_2}$ double layers has important effect on the in-plane piezoelectric polarization of  $\mathrm{MSi_2N_4}$ (M=Mo and W) monolayers.
Finally, we investigate the intrinsic piezoelectricity of  monolayer $\alpha_1$- and $\alpha_2$-  $\mathrm{MA_2Z_4}$ (M=Cr, Mo and W; A=Si and Ge; Z=N and P) expect  $\mathrm{CrGe_2N_4}$.  It is found that the  $\mathrm{MA_2P_4}$ have more stronger  piezoelectricity than  $\mathrm{MA_2N_4}$. So,
 experimentally synthesizing monolayer $\mathrm{MA_2Z_4}$ containing P atoms  is  very promising for energy harvesting and piezoelectric sensing.

The rest of the paper is organized as follows. In the next
section, we shall give our computational details and methods  about piezoelectric coefficients.
 In the third section, we perform symmetry analysis for elastic and piezoelectric coefficients of $\alpha_i$- ($i$=1 to 6) $\mathrm{MA_2Z_4}$. In the fourth  sections, we shall present main results and analysis. Finally, we shall give our  conclusions in the fifth section.
\begin{table}
\centering \caption{The optimized lattice constants of $\alpha_i$- ($i$=1 to 6) $\mathrm{MSi_2N_4}$ (M=Mo and W)  using GGA ($\mathrm{{\AA}}$). }\label{tab0}
  \begin{tabular*}{0.48\textwidth}{@{\extracolsep{\fill}}ccccccc}
  \hline\hline
Name & $\alpha_1$ &  $\alpha_2$ & $\alpha_3$ & $\alpha_4$ & $\alpha_5$ & $\alpha_6$ \\\hline\hline
 $\mathrm{MoSi_2N_4}$   &2.91&  2.90 &  2.84 & 2.84 & 2.86 & 2.85\\
 $\mathrm{WSi_2N_4}$   &2.91 & 2.90  & 2.84  & 2.84 & 2.87 & 2.85\\ \hline\hline
\end{tabular*}
\end{table}

\section{Computational detail}
Based on  the density functional theory (DFT)\cite{1}, our  simulations are carried out   as implemented
in the plane-wave code VASP\cite{pv1,pv2,pv3}. The exchange-correlation functional is treated within  popular generalized gradient
approximation of Perdew, Burke and  Ernzerhof  (GGA-PBE)\cite{pbe}  to perform the structural relaxation and the calculations of the elastic and
piezoelectric tensors.  For energy band calculations, the spin orbital coupling (SOC)
is also taken into account due to containing early transition metal.
Projector-augmented wave pseudopotentials are used with a cutoff energy of 500 eV for plane-wave expansions.
A vacuum spacing of more than 20 $\mathrm{{\AA}}$ is adopted to prevent any interactions
between the adjacent periodic images of the 2D monolayers. The total energy  convergence criterion is set
to $10^{-8}$ eV, and  the
atomic positions  are optimized until all components of
the forces on each atom are reduced to values  less than 0.0001 $\mathrm{eV.{\AA}^{-1}}$.
We calculate the coefficients of  elastic stiffness tensor  $C_{ij}$  by using strain-stress relationship (SSR)  and   the piezoelectric stress coefficients $e_{ij}$  by DFPT method\cite{pv6}.
A Monkhorst-Pack mesh of 15$\times$15$\times$1  in the first Brillouin zone
is sampled   for $C_{ij}$, and  9$\times$16$\times$1 for $e_{ij}$.
The 2D elastic coefficients $C^{2D}_{ij}$
 and   piezoelectric stress coefficients $e^{2D}_{ij}$
have been renormalized by the the length of unit cell along z direction ($Lz$):  $C^{2D}_{ij}$=$Lz$$C^{3D}_{ij}$ and $e^{2D}_{ij}$=$Lz$$e^{3D}_{ij}$.
However, the $d_{ij}$ is independent of $Lz$.

\begin{figure*}
  \includegraphics[width=12cm]{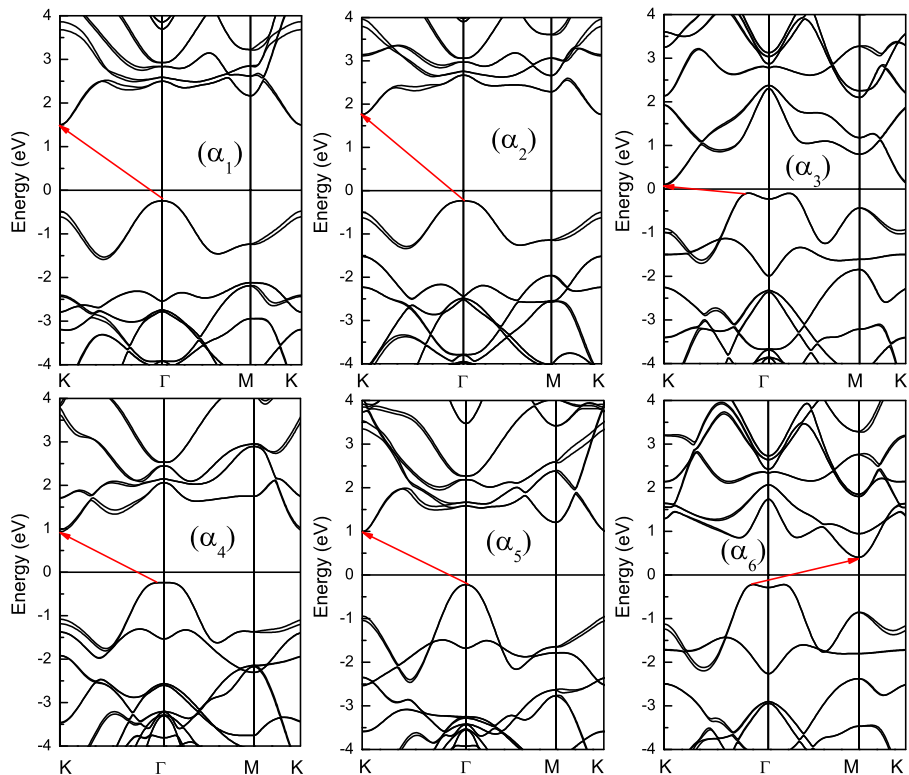}
  \caption{(Color online) The energy band structures of $\alpha_i$- ($i$=1 to 6) $\mathrm{MoSi_2N_4}$ using GGA+SOC, and the VBM and CBM are connected by red arrow.  }\label{band}
\end{figure*}

\begin{figure}
  \includegraphics[width=7cm]{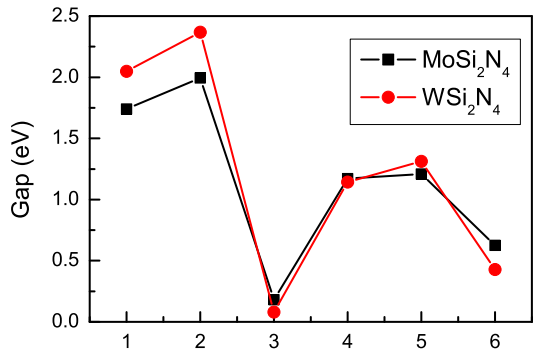}
  \caption{(Color online)The energy band gaps of $\alpha_i$- ($i$=1 to 6) $\mathrm{MoSi_2N_4}$ and $\mathrm{WSi_2N_4}$  using GGA+SOC.}\label{band1}
\end{figure}

\begin{figure}
  \includegraphics[width=7cm]{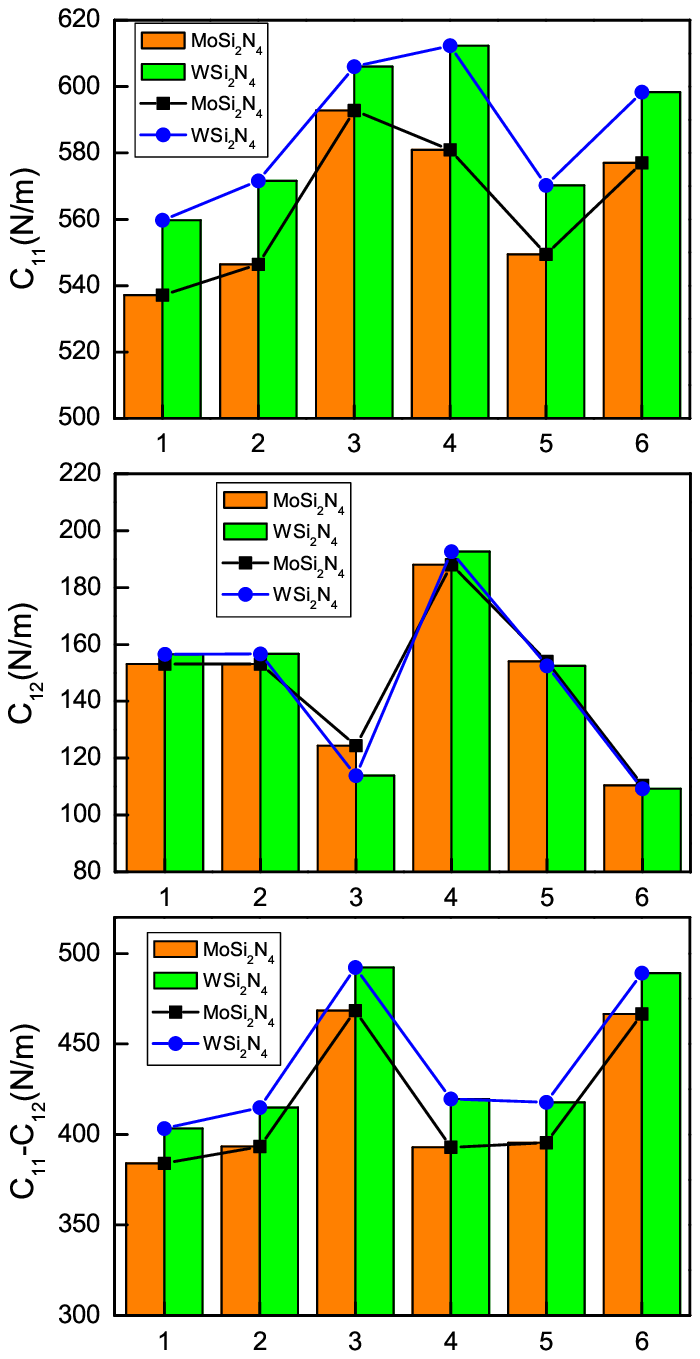}
  \caption{(Color online) The elastic constants  $C_{ij}$ of $\alpha_i$- ($i$=1 to 6) $\mathrm{MoSi_2N_4}$ and $\mathrm{WSi_2N_4}$.}\label{c}
\end{figure}
\begin{figure}
  \includegraphics[width=7cm]{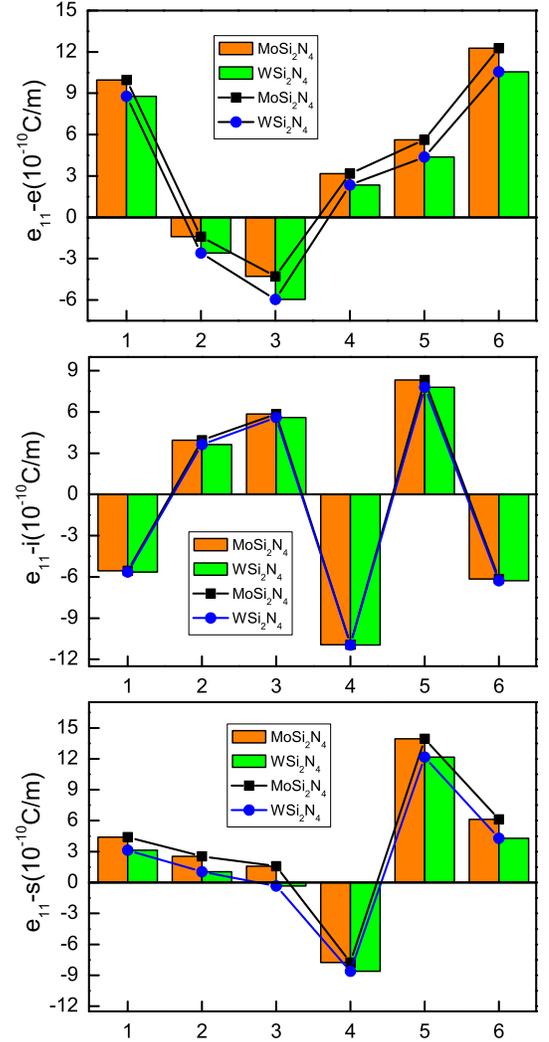}
  \caption{(Color online) The  piezoelectric stress coefficients  $e_{11}$,  the ionic contribution and electronic contribution to $e_{11}$ of $\alpha_i$- ($i$=1 to 6) $\mathrm{MoSi_2N_4}$ and $\mathrm{WSi_2N_4}$.  }\label{e}
\end{figure}

\begin{table}
\centering \caption{The $d_{11}$ of $\alpha_i$- ($i$=1 to 6) $\mathrm{MSi_2N_4}$ (M=Mo and W)  using GGA (pm/V). }\label{tab0-1}
  \begin{tabular*}{0.48\textwidth}{@{\extracolsep{\fill}}ccccccc}
  \hline\hline
Name & $\alpha_1$ &  $\alpha_2$ & $\alpha_3$ & $\alpha_4$ & $\alpha_5$ & $\alpha_6$ \\\hline\hline
 $\mathrm{MoSi_2N_4}$   &1.15 &  0.65&  0.34 & -1.98 & 3.53 & 1.32\\
 $\mathrm{WSi_2N_4}$   &0.78 &0.25 & -0.07  & -2.05 & 2.91 & 0.88\\ \hline\hline
\end{tabular*}
\end{table}

\section{Symmetry Analysis}
 The piezoelectric  stress tensors  $e_{ijk}$ and strain tensor $d_{ijk}$  is defined as:
 \begin{equation}\label{pe0}
      e_{ijk}=\frac{\partial P_i}{\partial \varepsilon_{jk}}=e_{ijk}^{elc}+e_{ijk}^{ion}
 \end{equation}
and
 \begin{equation}\label{pe0-1}
   d_{ijk}=\frac{\partial P_i}{\partial \sigma_{jk}}=d_{ijk}^{elc}+d_{ijk}^{ion}
 \end{equation}
where  $P_i$, $\varepsilon_{jk}$ and $\sigma_{jk}$ are polarization vector, strain and stress, respectively.
The $e_{ijk}^{elc}$  ($d_{ijk}^{elc}$) is the clamped-ion
piezoelectric tensors resulting from the pure electronic contribution.  The relaxed-ion
piezoelectric tensors $e_{ijk}$ ($d_{ijk}$) is obtained from the sum of ionic
and electronic contributions. The $d_{ijk}$ can be connected with  $e_{ijk}$ by the elastic stiffness tensor  $C_{ijkl}$.
By employing the frequently used Voigt notation (11$\rightarrow$1,
22$\rightarrow$2, 33$\rightarrow$3, 23$\rightarrow$4, 31$\rightarrow$5 and 12$\rightarrow$6),
the elastic tensor  $C_{ijkl}$, piezoelectric tensors $e_{ijk}$ and $d_{ijk}$ become into $C_{ij}$ (6$\times$6  matrix), $e_{ij}$  (3$\times$6  matrix) and $d_{ij}$ (3$\times$6  matrix). The symmetry of crystal
structure will further reduce the number of independent  $C_{ij}$, $e_{ij}$ and $d_{ij}$ tensors.

By intercalating  $\mathrm{MoS_2}$-type  $\mathrm{MZ_2}$ monolayer into InSe-type  $\mathrm{A_2Z_2}$ monolayer,
six $\alpha_i$ and six $\beta_i$ ($i$=1 to 6)  $\mathrm{MA_2Z_4}$ monolayers can be constructed\cite{m20}.
The six $\alpha_i$ have the same $P\bar{6}m2$ space group due to inserting  2H-$\mathrm{MoS_2}$-type  $\mathrm{MZ_2}$ monolayer   into $\alpha$-InSe-type  $\mathrm{A_2Z_2}$ double layers, which break inversion symmetry. The six $\beta_i$ are built by intercalating 1T-$\mathrm{MoS_2}$-type  $\mathrm{MZ_2}$ monolayer into $\beta$-InSe-type  $\mathrm{A_2Z_2}$ double layers with the same $P\bar{3}m1$ space group, which keep inversion symmetry.
Therefore, $\alpha_i$-  ($i$=1 to 6)  $\mathrm{MA_2Z_4}$ monolayers are piezoelectric.

 The six  $\alpha_i$  geometric structures of the   $\mathrm{MA_2Z_4}$ monolayer are
plotted in \autoref{t0}. All considered six $\alpha_i$ crystal structures  have  the same $\bar{6}m2$ point group. Only  the in-plane  piezoelectric effect is  allowed in
monolayer $\alpha_i$- ($i$=1 to 6) $\mathrm{MA_2Z_4}$, when a uniaxial in-plane  strain is applied. For 2D semiconductors, in general,
in-plane  stresses and strains are only allowed,
while the out-of-plane is strain/stress free\cite{q9,q10,q11}. And then the  $e_{ij}$,  $d_{ij}$ and $C_{ij}$  can be written as:
 \begin{equation}\label{pe1}
  \left(
    \begin{array}{ccc}
      e_{11} &-e_{11} & 0 \\
    0 &0 & -e_{11}\\
      0 & 0 & 0 \\
    \end{array}
  \right)
  \end{equation}
  \begin{equation}\label{pe1}
  \left(
    \begin{array}{ccc}
        d_{11} & -d_{11} & 0 \\
    0 &0 & -2d_{11} \\
      0 & 0 & 0 \\
    \end{array}
  \right)
  \end{equation}
  \begin{equation}\label{pe1}
    \left(
    \begin{array}{ccc}
      C_{11} & C_{12} &0 \\
     C_{12} & C_{11} &0 \\
     0 & 0 & \frac{C_{11}-C_{12}}{2} \\
    \end{array}
  \right)
   \end{equation}
 The forms of these piezoelectric and stiffness constants  are the same as those for TMD monolayers\cite{q9,q11} due to the same  point group.
By  $e_{ik}$=$d_{ij}C_{jk}$, the only in-plane $d_{11}$ is found to be:
\begin{equation}\label{pe2-7}
    d_{11}=\frac{e_{11}}{C_{11}-C_{12}}
\end{equation}

\begin{table*}
\centering \caption{The optimized lattice constants of $\alpha_i$- ($i$=1 to 2)  $\mathrm{MA_2Z_4}$ (M=Cr, Mo and W; A=Si and Ge; Z=N and P) expect  $\mathrm{CrGe_2N_4}$  using GGA ($\mathrm{{\AA}}$). }\label{tab1}
  \begin{tabular*}{0.96\textwidth}{@{\extracolsep{\fill}}cccccccccccc}
  \hline\hline
Name & $\mathrm{CrSi_2N_4}$ &  $\mathrm{MoSi_2N_4}$ & $\mathrm{WSi_2N_4}$ & $\mathrm{MoGe_2N_4}$ & $\mathrm{WGe_2N_4}$ & $\mathrm{CrSi_2P_4}$&$\mathrm{MoSi_2P_4}$&$\mathrm{WSi_2P_4}$&$\mathrm{CrGe_2P_4}$&$\mathrm{MoGe_2P_4}$&$\mathrm{WGe_2P_4}$ \\\hline\hline
 $\alpha_1$ &2.84&  2.91 & 2.91 &  3.02 & 3.02&  3.42 & 3.47 & 3.47 &3.50 & 3.54 &3.54\\
  $\alpha_2$  &2.84 & 2.90 & 2.90  & 3.01  &3.01 & 3.41 & 3.45 & 3.46 &3.49 & 3.53 &  3.53\\ \hline\hline
\end{tabular*}
\end{table*}

\section{Main calculated results}
 Firstly,  we discuss the  structural relation among six $\alpha_i$  crystal structures of  $\mathrm{MA_2Z_4}$.
 According to the relative positions of M and A atoms, the six  $\alpha_i$  crystal structure  can be divided into three categories: A ($\alpha_1$, $\alpha_2$), B ($\alpha_3$, $\alpha_4$) and C ($\alpha_5$, $\alpha_6$).
The different categories can be connected by translation operation.  The $\alpha_3$  can be attained by  translating $\mathrm{A_2Z_2}$ double layers of   $\alpha_2$ along the green line of top view of $\alpha_2$ in \autoref{t0} with  the transfixion of  $\mathrm{MZ_2}$ monolayer.
 The $\alpha_6$  can be attained from  $\alpha_4$ by similar translation operation. The different structures in the same category can be related by mirror or rotation operations.  The  $\alpha_2$ can be built by mirroring $\mathrm{A_2Z_2}$ double layers of $\alpha_1$ with respect to the vertical  surface defined by two green lines of top and side views of $\alpha_1$. The $\alpha_4$ ($\alpha_6$) can be constructed by rotating the $\mathrm{A_2Z_2}$ double layers of $\alpha_3$ ($\alpha_5$) with $\pi$/3 along the vertical axis defined by linking two A atoms.

  \begin{figure}
  \includegraphics[width=7cm]{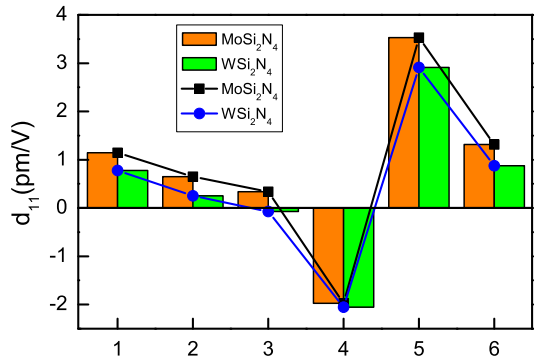}
  \caption{(Color online) The  piezoelectric strain coefficients  $d_{11}$ of $\alpha_i$- ($i$=1 to 6) $\mathrm{MoSi_2N_4}$ and $\mathrm{WSi_2N_4}$.  }\label{d}
\end{figure}

\begin{table*}
\centering \caption{The $d_{11}$ of $\alpha_i$- ($i$=1 to 2)  $\mathrm{MA_2Z_4}$ (M=Cr, Mo and W; A=Si and Ge; Z=N and P) expect  $\mathrm{CrGe_2N_4}$  using GGA (pm/V). }\label{tab1-1}
  \begin{tabular*}{0.96\textwidth}{@{\extracolsep{\fill}}cccccccccccc}
  \hline\hline
Name & $\mathrm{CrSi_2N_4}$ &  $\mathrm{MoSi_2N_4}$ & $\mathrm{WSi_2N_4}$ & $\mathrm{MoGe_2N_4}$ & $\mathrm{WGe_2N_4}$ & $\mathrm{CrSi_2P_4}$&$\mathrm{MoSi_2P_4}$&$\mathrm{WSi_2P_4}$&$\mathrm{CrGe_2P_4}$&$\mathrm{MoGe_2P_4}$&$\mathrm{WGe_2P_4}$ \\\hline\hline
 $\alpha_1$ &1.24&	1.15&	0.78&	1.85&	1.31&	6.03&	4.91&	4.16&	6.12&	5.27&	4.36\\
  $\alpha_2$  &1.42&	0.65&	0.25&	0.75&	0.26&	3.96&	2.64&	1.65&	5.06&	3.87&	2.77\\ \hline\hline
\end{tabular*}
\end{table*}

 It has been proved that monolayer  $\mathrm{MA_2Z_4}$ (M=Cr, Mo and W; A=Si and Ge; Z=N and P) expect  $\mathrm{CrGe_2N_4}$  with 34 VEC are non-magnetic with $\alpha_1$ or $\alpha_2$ crystal structure, and are both dynamically and thermodynamically\cite{m20}.  A piezoelectric material  should  be a semiconductor for
prohibiting current leakage. Only $\mathrm{MSi_2N_4}$ (M=Mo and W) monolayers are  semiconductors for all six $\alpha_i$ crystal structures.
So, we mainly study the structure  effect on intrinsic piezoelectricity of $\mathrm{MSi_2N_4}$ (M=Mo and W). The
structural parameters  of $\alpha_i$- ($i$=1 to 6)  $\mathrm{MSi_2N_4}$ (M=Mo and W) are  optimized, and the lattice constants are listed in \autoref{tab0}.
It is found that the lattice constants between $\mathrm{MoSi_2N_4}$ and $\mathrm{WSi_2N_4}$  with the same phase are almost the same.
The size of these  lattice constants can also be  classified into A, B and C, which declare that the relative positions of M and A atoms determine lattice constants.

Next, we use optimized crystal structures  to investigate their electronic structures.
 Although the SOC has little effects on the energy band gaps of  $\mathrm{MSi_2N_4}$ (M=Mo and W) monolayers,  the SOC  can produce  observed  spin-orbit splitting  in the valence bands   at K point\cite{m21}. Because the energy band outlines between $\mathrm{MoSi_2N_4}$ and $\mathrm{WSi_2N_4}$ are very similar, only the energy bands of $\alpha_i$- ($i$=1 to 6) $\mathrm{MoSi_2N_4}$ are shown in \autoref{band} using GGA+SOC. It is clearly seen that they all are indirect gap semiconductors, and the gap range is 0.18 eV to 1.99 eV.  The position  of  conduction band minimum (CBM) for all six $\alpha_i$  is  at K point, except for $\alpha_6$ at M point.  The valence band maximum (VBM) of  $\alpha_1$, $\alpha_2$  and $\alpha_5$ is  at $\Gamma$ point, while the one of  $\alpha_3$, $\alpha_4$  and $\alpha_6$ is slightly off $\Gamma$ point, and  at one point along the $\Gamma$-K line.
The energy band gaps of $\alpha_i$- ($i$=1 to 6) $\mathrm{MoSi_2N_4}$ and $\mathrm{WSi_2N_4}$  using GGA+SOC are plotted \autoref{band1}. It is clearly seen that the structural dependence of band gap of  $\mathrm{WSi_2N_4}$ is the same with one of  $\mathrm{MoSi_2N_4}$, and the gap ranges from 0.08 eV to 2.37 eV. Therefore, it is very effective to tune the electronic structures  of  $\mathrm{MSi_2N_4}$ (M=Mo and W) monolayers by  translating or rotating  $\mathrm{Si_2N_2}$ bilayer.
\begin{figure}
  \includegraphics[width=7cm]{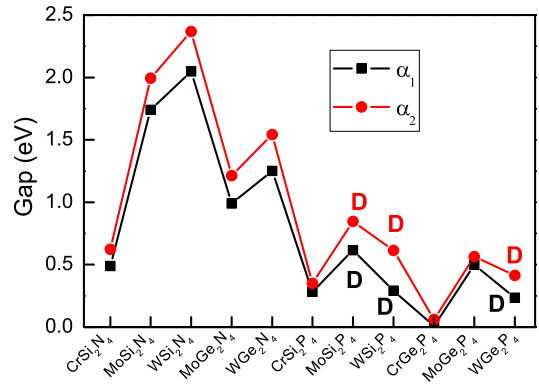}
  \caption{(Color online)The energy band gaps of  $\alpha_i$- ($i$=1 to 2)  $\mathrm{MA_2Z_4}$ (M=Cr, Mo and W; A=Si and Ge; Z=N and P) expect  $\mathrm{CrGe_2N_4}$  using GGA+SOC. The direct band gap is marked by "D", and the unmarked one is indirect band gap.}\label{band2}
\end{figure}

To calculate the $d_{11}$,  two independent elastic stiffness coefficients  ($C_{11}$ and $C_{12}$) of  $\alpha_i$- ($i$=1 to 6) $\mathrm{MoSi_2N_4}$ and $\mathrm{WSi_2N_4}$ are attained by SSR, which are plotted in \autoref{c}, together with $C_{11}$-$C_{12}$.  For six structures,  all calculated elastic coefficients of $\mathrm{MoSi_2N_4}$ and $\mathrm{WSi_2N_4}$ satisfy the Born stability criteria\cite{ela}, which means that they all are mechanically stable. Similar structural dependence of $C_{11}$, $C_{12}$ and $C_{11}$-$C_{12}$ can be observed between $\mathrm{MoSi_2N_4}$ and $\mathrm{WSi_2N_4}$.
It is found that the  $C_{12}$  of two structures in the same category are very close, if the two structures are connected by mirror operation  ($\alpha_1$ and  $\alpha_2$).  However, the  $C_{12}$  has obvious difference for two structures related by rotation operation  ($\alpha_3$  and  $\alpha_4$  or $\alpha_5$  and  $\alpha_6$). The  $\alpha_4$ ($\alpha_5$)  has the larger
 $C_{12}$ than $\alpha_3$ ($\alpha_6$). Between  $\alpha_4$ ($\alpha_3$) and $\alpha_5$ ($\alpha_6$), the difference is only the position of Si atom.
It is found that the $C_{11}$-$C_{12}$  of $\alpha_1$, $\alpha_2$,   $\alpha_4$ and  $\alpha_5$ are close, and  $\alpha_3$ and  $\alpha_6$ have the larger $C_{11}$-$C_{12}$, which is against the $d_{11}$ according to \autoref{pe2-7}. These elastic stiffness coefficients are very larger than ones of
 TMD monolayers\cite{q9,q11},  which  indicates that
$\mathrm{MoSi_2N_4}$ and $\mathrm{WSi_2N_4}$ are not easy to be deformed.
\begin{figure}
  \includegraphics[width=8cm]{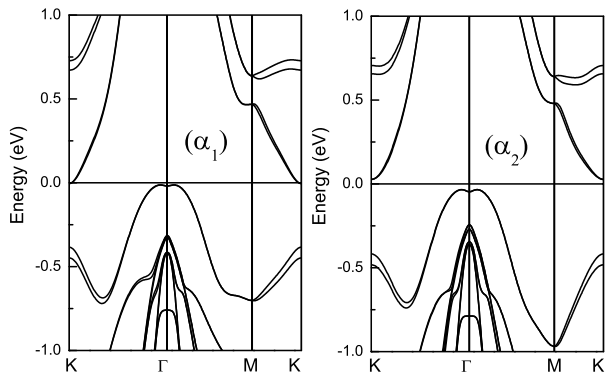}
  \caption{The energy band structures of  $\alpha_i$- ($i$=1 to 2)   $\mathrm{CrGe_2P_4}$  using GGA+SOC.}\label{band-c}
\end{figure}

 Another key physical quantity $e_{11}$ of   $\alpha_i$- ($i$=1 to 6) $\mathrm{MoSi_2N_4}$ and $\mathrm{WSi_2N_4}$ are calculated to attain $d_{11}$. Their   piezoelectric coefficients  $e_{11}$ along with the ionic contribution and electronic contribution to $e_{11}$  are shown \autoref{e}.
 It is clearly seen that the similar structural dependence between  $\mathrm{MoSi_2N_4}$ and $\mathrm{WSi_2N_4}$ can be observed.
 It is found that the ionic contribution of two structures  connected by mirror or rotation operations  in the same category  has opposite sign.  In the different category, the ionic contribution of two structures  connected by translation operation   has the same sign.
 Calculated results show that the ionic contribution and electronic contribution  have opposite sign for all $\alpha_i$ except $\alpha_5$.
  For A and B categories, the electronic contribution has similar structural dependence  with  ionic contribution.
  In the C category, the rotation operation gives rise to the identical signs for electronic contribution from $\alpha_6$ to $\alpha_5$.
In considered six structures, the $e_{11}$ of $\alpha_5$ has the largest value, which is due to superposed  ionic contribution and electronic contribution. The  $e_{11}$ with  $\alpha_5$ phase is 13.95$\times$$10^{-10}$ C/m for $\mathrm{MoSi_2N_4}$,  and	12.17$\times$$10^{-10}$ C/m for $\mathrm{WSi_2N_4}$.  These $e_{11}$ are very larger than ones of 2D TMD, metal oxides,  III-V
semiconductor and Janus TMD materials\cite{q7,q9,q11}.
The  $e_{11}$  of  experimentally synthesized  $\alpha_1$-$\mathrm{MoSi_2N_4}$ and $\mathrm{WSi_2N_4}$ is 4.40$\times$$10^{-10}$ C/m  and 3.14$\times$$10^{-10}$ C/m, which
are  comparable to that of most  2D materials, such as  TMD and Janus TMD materials\cite{q7,q9,q11}.
Using the calculated $C_{11}$-$C_{12}$ and $e_{11}$, the $d_{11}$ can be attained according to \autoref{pe2-7}, which are shown in \autoref{d}. From  $\alpha_1$ to  $\alpha_6$, the $d_{11}$ and $e_{11}$ show very analogical structural dependence. For $\alpha_5$ phase,  the  $d_{11}$ has the largest value of  3.53 pm/V for $\mathrm{MoSi_2N_4}$,  and	2.91 pm/V for $\mathrm{WSi_2N_4}$.
 For experimentally synthesized  $\alpha_1$-$\mathrm{MoSi_2N_4}$ and $\mathrm{WSi_2N_4}$,  the  $d_{11}$  is 1.14 pm/V  and 0.78 pm/V, which are  smaller than that of 2D TMD\cite{q9,q11}  due to very large $C_{11}$-$C_{12}$. The related $d_{11}$ are listed in \autoref{tab0-1}.
\begin{figure}
  \includegraphics[width=7cm]{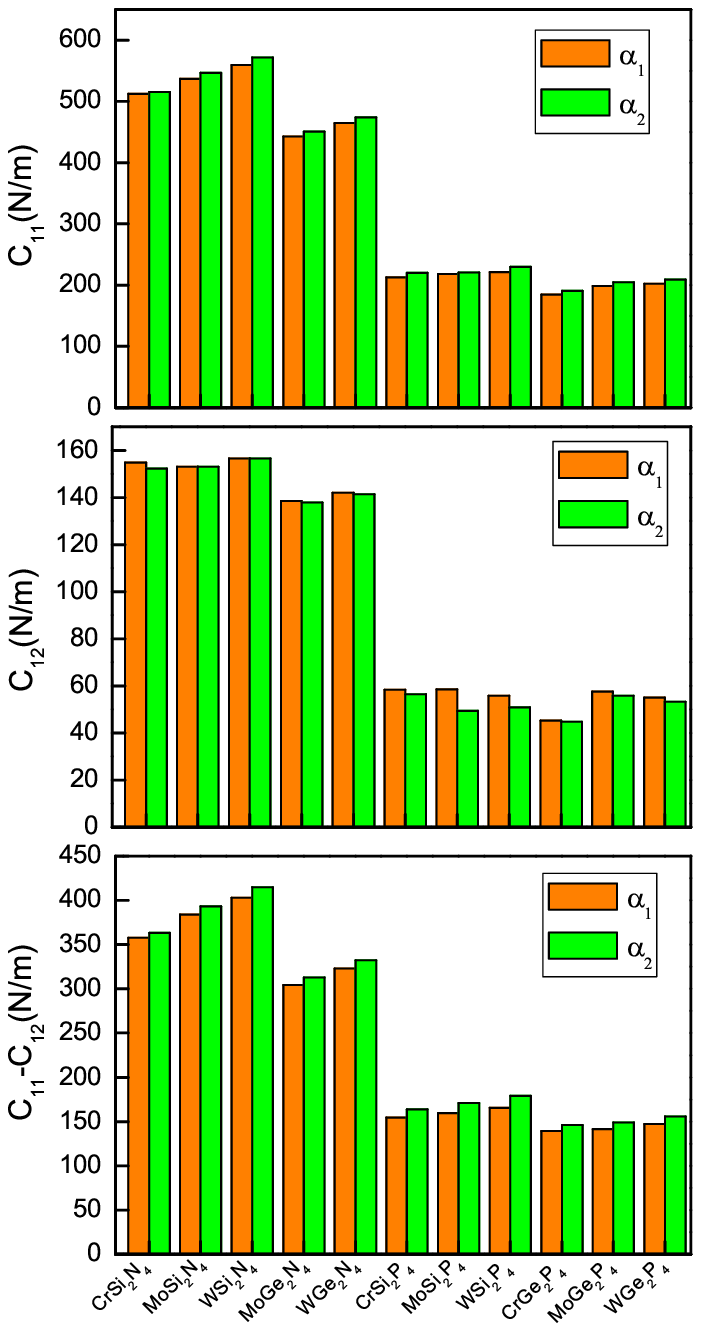}
  \caption{(Color online) The elastic constants  $C_{ij}$ of $\alpha_i$- ($i$=1 to 2)  $\mathrm{MA_2Z_4}$ (M=Cr, Mo and W; A=Si and Ge; Z=N and P) expect  $\mathrm{CrGe_2N_4}$.}\label{c1}
\end{figure}

For  $\alpha_1$ and  $\alpha_2$ phases, the  monolayer  $\mathrm{MA_2Z_4}$ (M=Cr, Mo and W; A=Si and Ge; Z=N and P) expect  $\mathrm{CrGe_2N_4}$ are all semiconductors using GGA+SOC. The energy band gaps of  $\alpha_i$- ($i$=1 to 2)  $\mathrm{MA_2Z_4}$ (M=Cr, Mo and W; A=Si and Ge; Z=N and P) expect  $\mathrm{CrGe_2N_4}$  using GGA+SOC are plotted in \autoref{band2}. It is found that the gap of  $\mathrm{CrGe_2P_4}$ is very small, and  0.008 eV for  $\alpha_1$ phase and 0.061 eV for $\alpha_2$ phase. To unambiguously indicate them to be semiconductors,  the energy band structures of  $\alpha_i$- ($i$=1 to 2)  $\mathrm{CrGe_2P_4}$  using GGA+SOC are shown in \autoref{band-c}.
 For the same material, the gap with $\alpha_2$ phase is larger than one of $\alpha_1$ phase.  For $\mathrm{MA_2N_4}$, the gap increases with M from Cr to Mo to W, while  the gap of  $\mathrm{MA_2P_4}$ firstly increases, and then decreases.  Another reason is that their  enthalpies of formation between  $\alpha_1$ and  $\alpha_2$ phases are very close\cite{m20}.
So, we investigate the intrinsic piezoelectricity of the 11 kinds of materials with $\alpha_1$ and  $\alpha_2$ phases.
The optimize lattice constants are listed in \autoref{tab1}, and the lattice constants between $\alpha_1$ and  $\alpha_2$ phases for the same material
almost the same, which is because the $\alpha_1$ and  $\alpha_2$ phases are in the same A class.
With element changing from Cr to Mo to W, from Si to Ge, and from N to P, the lattice constants of both $\alpha_1$ and $\alpha_2$ phases increase, which is due to increasing atomic radius.

\begin{figure}
  \includegraphics[width=7cm]{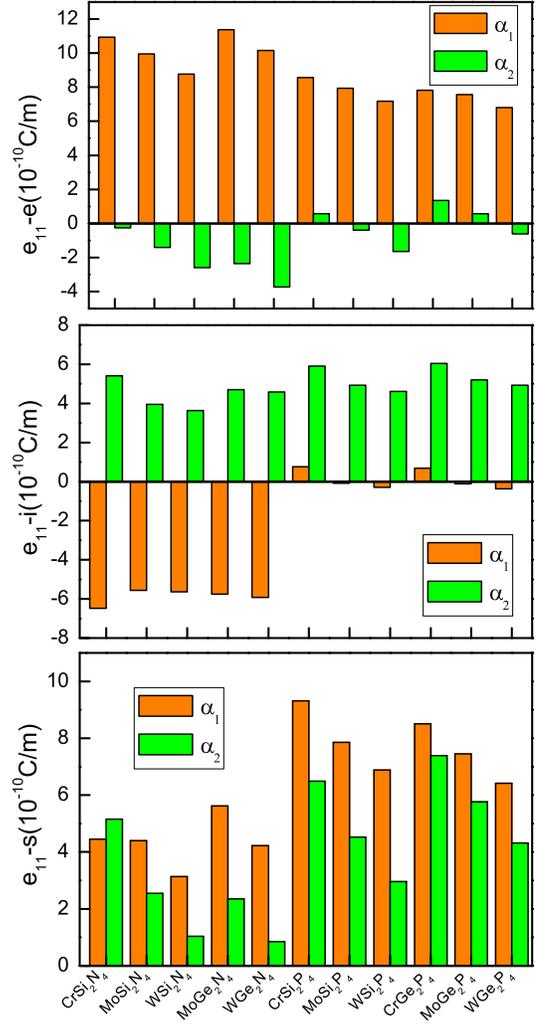}
  \caption{(Color online) The  piezoelectric stress coefficients  $e_{11}$,  the ionic contribution and electronic contribution to $e_{11}$ of $\alpha_i$- ($i$=1 to 2)  $\mathrm{MA_2Z_4}$ (M=Cr, Mo and W; A=Si and Ge; Z=N and P) expect  $\mathrm{CrGe_2N_4}$.  }\label{e1}
\end{figure}

The elastic constants  $C_{ij}$ of $\alpha_i$- ($i$=1 to 2)  $\mathrm{MA_2Z_4}$ (M=Cr, Mo and W; A=Si and Ge; Z=N and P) expect  $\mathrm{CrGe_2N_4}$ are plotted in \autoref{c1}.  For all studied materials, the $C_{11}$-$C_{12}$  with  $\alpha_2$ phase are larger than ones of  $\alpha_1$ phase, which is due to larger $C_{11}$ and smaller $C_{12}$. The  $C_{ij}$ of $\alpha_i$- ($i$=1 to 2)  $\mathrm{MA_2Z_4}$ containing P atom are very smaller than ones including N atom, which is favor of $d_{11}$. The  piezoelectric stress coefficients  $e_{11}$  of $\alpha_i$- ($i$=1 to 2)  $\mathrm{MA_2Z_4}$ (M=Cr, Mo and W; A=Si and Ge; Z=N and P) expect  $\mathrm{CrGe_2N_4}$ together with  the ionic contribution and electronic contribution to $e_{11}$ are shown in \autoref{e1}. When M changes from Cr to Mo to W with the same A and Z atoms,   the electronic contribution of  $\mathrm{MA_2Z_4}$ with  $\alpha_1$ phase decreases, while the one of $\alpha_2$ phase changes toward more negative value. It is found that the electronic contribution (absolute value) of all  $\mathrm{MA_2Z_4}$ with  $\alpha_2$ phase is smaller than one of  $\alpha_1$.  The ionic contribution of all materials with $\alpha_2$ phase is positive, which is the same with the electronic contribution of $\alpha_1$.
 With M from Cr to Mo to W,   the ionic contribution of  $\mathrm{MA_2Z_4}$ of   $\alpha_2$ phase with the same A and Z atoms  decreases, while the one of $\alpha_1$ phase changes toward more negative value except for  $\mathrm{CrSi_2N_4}$. It is clearly seen that the $e_{11}$ of all materials are positive values. The $e_{11}$ of  $\mathrm{MA_2Z_4}$ containing P atom with the same M and A atoms is larger than one including N atom for both  $\alpha_1$ and  $\alpha_2$ phases.  For  $\alpha_1$ phase, the $e_{11}$ ranges from 3.14$\times$$10^{-10}$ C/m to 9.31$\times$$10^{-10}$ C/m, and the whole range for $\alpha_2$ phase is 0.85$\times$$10^{-10}$ C/m  to 7.39$\times$$10^{-10}$ C/m.
  \begin{figure}
  \includegraphics[width=7cm]{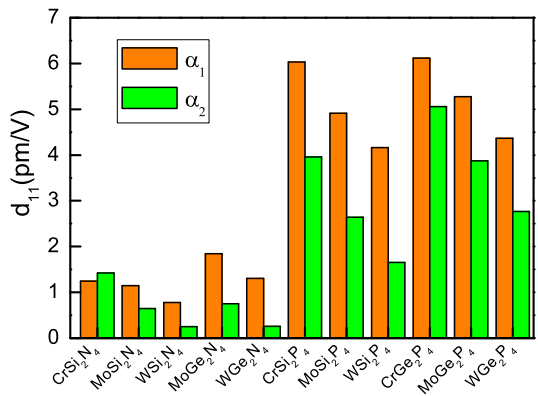}
  \caption{(Color online) The  piezoelectric strain coefficients  $d_{11}$ of  $\alpha_i$- ($i$=1 to 2)  $\mathrm{MA_2Z_4}$ (M=Cr, Mo and W; A=Si and Ge; Z=N and P) expect  $\mathrm{CrGe_2N_4}$.  }\label{d1}
\end{figure}

 Finally, the  piezoelectric strain coefficients  $d_{11}$ of  $\alpha_i$- ($i$=1 to 2)  $\mathrm{MA_2Z_4}$ (M=Cr, Mo and W; A=Si and Ge; Z=N and P) expect  $\mathrm{CrGe_2N_4}$ are plotted in \autoref{d1}.  The related $d_{11}$ are also summarized in \autoref{tab1-1}. For  $\alpha_1$ phase, the range of  $d_{11}$ is 0.78 pm/V to 6.12 pm/V, and the  range changes from 0.25 pm/V to 5.06 pm/V  for $\alpha_2$ phase. The change trend of  $d_{11}$  as a function of material  is very similar with one of $e_{11}$. It is clearly seen that monolayer $\mathrm{MA_2Z_4}$ containing P atom have more excellent piezoelectric response due to high $d_{11}$.
 The most  $d_{11}$ of them  are larger than $d_{33}$ = 3.1 pm/V of  familiar bulk piezoelectric  wurtzite GaN\cite{zh1}.
So, it is highly recommended to synthesize monolayer $\mathrm{MA_2Z_4}$ containing P atom, such as  $\alpha_1$-$\mathrm{CrSi_2P_4}$, $\alpha_1$-$\mathrm{MoSi_2P_4}$, $\alpha_1$-$\mathrm{CrGe_2P_4}$, $\alpha_1$-$\mathrm{MoGe_2P_4}$ and $\alpha_2$-$\mathrm{CrGe_2P_4}$.

\section{Conclusion}
We have demonstrated strong structure  effect on intrinsic piezoelectricity in septuple-atomic-layer  $\mathrm{MSi_2N_4}$ (M=Mo and W)
through first-principles simulations.   The same structural dependence on $d_{11}$ and $e_{11}$, together with the  ionic and  electronic contributions to  $e_{11}$ between $\mathrm{MoSi_2N_4}$ and  $\mathrm{WSi_2N_4}$ monolayers is found, and the  $\alpha_5$  phase  has  large piezoelectric coefficients. The intrinsic piezoelectricity of  monolayer  $\mathrm{MA_2Z_4}$ (M=Cr, Mo and W; A=Si and Ge; Z=N and P) with $\alpha_1$ and $\alpha_2$ phases expect  $\mathrm{CrGe_2N_4}$ are explored, and the  monolayer $\mathrm{MA_2P_4}$  have more stronger  piezoelectric polarization than  monolayer $\mathrm{MA_2Z_4}$  including N atom.
The largest $d_{11}$  among $\mathrm{MA_2N_4}$ materials  only is 1.85 pm/V, and  the largest $d_{11}$ of $\mathrm{MA_2P_4}$ is up to 6.12 pm/V.
 Among the studied 22 materials, the $d_{11}$ of monolayer  $\alpha_1$-$\mathrm{CrSi_2P_4}$, $\alpha_1$-$\mathrm{MoSi_2P_4}$, $\alpha_1$-$\mathrm{CrGe_2P_4}$, $\alpha_1$-$\mathrm{MoGe_2P_4}$ and $\alpha_2$-$\mathrm{CrGe_2P_4}$ are  greater than or close to 5 pm/V.
 These $d_{11}$ of  $\mathrm{MA_2P_4}$ compare favorably with piezoelectric coefficients of
familiar bulk piezoelectrics such as $\alpha$-quartz ($d_{11}$ = 2.3
pm/V), wurtzite GaN ($d_{33}$ = 3.1 pm/V) and wurtzite AlN ($d_{33}$ = 5.1 pm/V)\cite{zh1,zh2}.
Our works provide valuable guidance for
experimental synthesis efforts, and hope our study will stimulate more
research interest into $\mathrm{MA_2Z_4}$ family, especially for
its applications in piezoelectric field.

\section{Data availability}
The data that support the findings of this study are available from the corresponding author upon reasonable request.

\begin{acknowledgments}
This work is supported by the Natural Science Foundation of Shaanxi Provincial Department of Education (19JK0809). We are grateful to the Advanced Analysis and Computation Center of China University of Mining and Technology (CUMT) for the award of CPU hours and WIEN2k/VASP software to accomplish this work.
\end{acknowledgments}

\end{document}